\pdfoutput=1
\documentclass[3p]{elsarticle}

\usepackage{hyperref}

\usepackage{natbib}
\usepackage{cite}
\usepackage{amsmath,amssymb,amsfonts}
\usepackage{algorithmic}
\usepackage{graphicx}
\usepackage{textcomp}
\usepackage{xcolor}

\usepackage{tabularx,booktabs}
\usepackage{url}

\newcommand{\argmax}{\mathop{\mathrm{argmax\,}}}

\newcommand{\boldA}{{\boldsymbol{A}}}

\newcommand{\boldX}{{\boldsymbol{X}}}

\newcommand{\boldb}{{\boldsymbol{b}}}

\newcommand{\boldq}{{\boldsymbol{q}}}

\newcommand{\bolds}{{\boldsymbol{s}}}
\newcommand{\boldt}{{\boldsymbol{t}}}

\newcommand{\boldx}{{\boldsymbol{x}}}

\newcommand{\boldtheta}{{\boldsymbol{\theta}}}

\usepackage[ruled, algo2e, vlined]{algorithm2e}
\usepackage{amsmath,amssymb,amsfonts,amsthm,mathtools}
\usepackage{bbm}
\usepackage{lscape}

\newcommand{\bft}{\fontseries{b}\selectfont}

\journal{Journal of Informetrics}

\begin{document}

\begin{frontmatter}

\title{Poincare: Recommending Publication Venues via Treatment Effect Estimation}

\author[kyoto,aip]{Ryoma Sato\corref{mycorrespondingauthor}}
\ead{r.sato@ml.ist.i.kyoto-u.ac.jp}
\cortext[mycorrespondingauthor]{Corresponding author}

\author[kyoto,aip]{Makoto Yamada}
\ead{myamada@i.kyoto-u.ac.jp}

\author[kyoto,aip]{Hisashi Kashima}
\ead{kashima@i.kyoto-u.ac.jp}

\address[kyoto]{Kyoto University}
\address[aip]{RIKEN AIP}

\begin{abstract}
Choosing a publication venue for an academic paper is a crucial step in the research process. However, in many cases, decisions are based solely on the experience of researchers, which often leads to suboptimal results. Although there exist venue recommender systems for academic papers, they recommend venues where the paper is expected to be published. In this study, we aim to recommend publication venues from a different perspective. We estimate the number of citations a paper will receive if the paper is published in each venue and recommend the venue where the paper has the most potential impact. However, there are two challenges to this task. First, a paper is published in only one venue, and thus, we cannot observe the number of citations the paper would receive if the paper were published in another venue. Secondly, the contents of a paper and the publication venue are not statistically independent; that is, there exist selection biases in choosing publication venues. In this paper, we formulate the venue recommendation problem as a treatment effect estimation problem. We use a bias correction method to estimate the potential impact of choosing a publication venue effectively and to recommend venues based on the potential impact of papers in each venue. We highlight the effectiveness of our method using paper data from computer science conferences.
\end{abstract}


\end{frontmatter}

\section{Introduction}

Selecting where to publish an academic paper is crucial to ensure that the research becomes widely known in the research community. If an inappropriate venue is selected, the dedicated efforts for that study could end in vain. Traag \citep{traag2021inferring} found that high impact journals accelerated the citation speeds of the accepted papers. Xiao et al. \citep{xiao2020discovering} found that influential researchers employed strategic behaviors when selecting publication venues. However, in many cases, a publication venue is decided based on the experience of researchers. For example, researchers publish their works in certain venues just because they published their previous works there, and they rarely consider publishing their new papers in unfamiliar venues. It is desirable to take other venues into consideration so that the research will be widely known. However, it is difficult to investigate all choices of publication venues due to many options. For example, as many as $50$ venues are listed in the ``Database/Data Mining/Content Retrieval'' category in the AMiner conference ranking\footnote{\url{https://www.aminer.org/ranks/conf}}. The list of top conferences in the ``Machine Learning, Data Mining \& Artificial intelligence'' category in the Guide2Research\footnote{\url{https://www.guide2research.com/topconf/}} contains as many as $100$ venues. The number of options increases if we take short papers and workshops into account. Furthermore, it is difficult for the researchers to expect what will happen when they publish papers in unfamiliar venues.

Recently, data-driven approaches have been proposed to help researchers select publication venues \citep{yang2012venue, schnabel2016recommendations}. However, they just recommend venues where the paper is likely to be published. We call such venue recommender systems \emph{association-based recommender systems}. For example, assume that researcher X always publishes her work in conference Y. In this case, an association-based recommender system learns to recommend conference Y for works of researcher X. However, the initial choice of researcher X may be suboptimal, and these works would have possibly attracted more researchers in other communities if researcher X had published these works in other venues.

In this paper, we propose to recommend publication venues wherein the paper attracts many researchers. We measure the influence of a paper by the number of citations. The proposed method helps researchers maximize the influence of their research and the number of citations received by their papers. Besides, our approach is beneficial for not only authors but also reviewers of publication venues. Suppose a seemingly out-of-scope but interesting paper is submitted to a conference, and the reviewers are wondering if they should accept it or not. An association-based recommender system does not suggest acceptance just because that paper is unfamiliar in that venue. However, the paper may offer a novel perspective and attract many researchers in that research community. In contrast to an association-based recommender system, our proposed approach can point out the possibility that the seemingly out-of-scope paper attracts many audiences in the field by showing the potential number of citations in this venue, and the reviewers will be able to take this aspect into consideration.

\begin{figure*}[t]
  \centering
\includegraphics[width=\hsize]{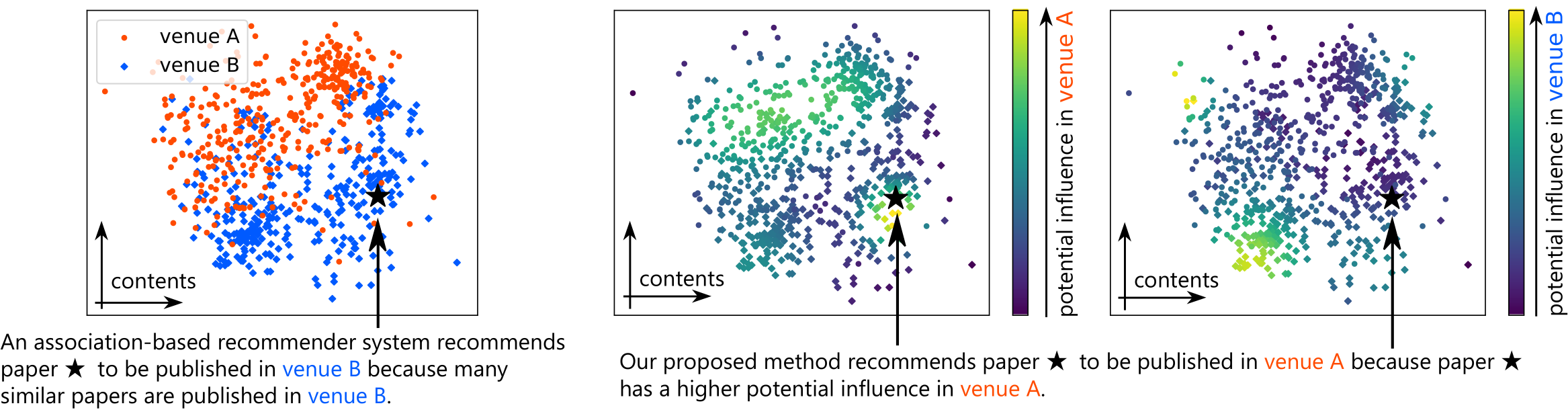}
\caption{Illustration (synthetic data): Each point represents a research paper, and the position indicates the contents (e.g., fields of study) of the paper. Association-based recommender systems recommend venues based on the likelihood of the venue. In contrast, our proposed method \textsc{Poincare} recommends venues based on potential influence.  }
 \label{fig: illust}
\end{figure*}

However, there are two challenges to achieving this goal. First, the same paper is not published in more than one venue, and therefore, we cannot observe how many citations the paper would receive if the paper were published in another venue. This dilemma is known as the \emph{fundamental problem of causal inference} \citep{holland1986statistics}. We need to estimate the number of citations in counterfactual venues. Secondly, the content of a paper and the venue in which that paper is published are not statistically independent. In other words, the assignment of a publication venue to a paper is not a randomized controlled trial. This selection bias makes the estimation of potential impact difficult.

In this paper, we propose \textsc{Poincare} (\underbar{Po}tential \underbar{in}fluence-based a\underbar{ca}demic publication venue \underbar{re}commendation), which uses a bias correction method to estimate the treatment effect of choosing a publication venue effectively and recommends the publication venue that has the highest potential effect. The major difference between association-based recommender systems and \textsc{Poincare} is depicted in Figure \ref{fig: illust}. The contributions are summarized as follows:

\begin{itemize}
    \item We formulate the venue recommendation problem through the lens of treatment effect estimation and advocate a new problem called the citation-aware publication venue recommendation problem, where we recommend a publication venue to maximize the influence of a paper.
    \item We propose to use a debiasing method to estimate the potential impact that a paper will have in a publication venue effectively.
    \item We empirically investigate the effect of our proposed method using a paper dataset from computer science conferences and illustrative simulated data.
\end{itemize}

The code and dataset are available at \url{https://github.com/joisino/poincare}.

\section{Problem Formulation} \label{sec: formulation}

We propose a new formulation for venue recommendation. This reformulation forms the core of our contribution. In the publication venue recommendation problem, we take a paper as input and recommend a publication venue. A paper $i$ is represented by a feature vector $\boldx_i \in \mathbb{R}^d$. For example, we use a bag of the fields of study as a feature vector in the experiments. Let $\mathcal{T}$ be the set of publication venues. Then, the publication venue recommendation problem is formalized as estimating function $r\colon \mathbb{R}^d \to \mathcal{T}$ that takes a feature of a paper as input and outputs a recommended venue. In the association-based formulation, a training dataset $\{(\boldx_i, t_i) \in \mathbb{R}^{d} \times \mathcal{T}\}_{i = 1, \dots, n}$ indicates that paper $i$ has a feature $\boldx_i$ and was published in venue $t_i$. Association-based methods recommend the most likely venue based on this data. In contrast, we use additional information to construct a recommender system. Specifically, a training dataset is represented by $\mathcal{D} = \{(\boldx_i, t_i, y_i) \in \mathbb{R}^{d} \times \mathcal{T} \times \mathbb{R}_+\}_{i = 1, \dots, n}$, where $(\boldx_i, t_i, y_i)$ indicates that paper $i$ has a feature $\boldx_i$, was published in venue $t_i$, and received $y_i$ citations in a certain period (e.g., $5$ years). We utilize this information about the number of citations for the venue recommendation. Specifically, we recommend a venue that maximizes the number of citations if that paper is published in the venue. Therefore, the problem we tackle in this paper is formalized as follows.

\vspace{0.1in}
\noindent \textbf{Citation-aware Publication Venue Recommendation Problem.} \newline
\textbf{Input:} A set of papers $\mathcal{D} = \{(\boldx_i, t_i, y_i) \in \mathbb{R}^{d} \times \mathcal{T} \times \mathbb{R}_+\}_{i = 1, \dots, n}$ \newline
\textbf{Output:} A recommender system $r\colon \mathbb{R}^d \to \mathcal{T}$ that recommends a venue that maximizes the number of citations.

A major challenge in this problem is the \emph{fundamental problem of causal inference}. For each paper $i$, we can know the number $y_i$ of citations if the paper is published in venue $t_i$, but we cannot know the number of citations if the paper is published in venue $t ~(\not= t_i)$ because each paper can be published in at most one venue. To tackle this issue, we utilize a treatment effect estimation method.

\vspace{0.1in}
\noindent \textbf{Rubin's potential outcome framework.} Rubin's potential outcome framework is one of the standard frameworks for causal inference, developed by Rubin \citep{rubin1974estimating, rubin1978bayesian, holland1986statistics, imbens2015causal}. The other standard approach for causal inference is Pearl's model \citep{pearl1995causal, pearl2009causality, pearl2009causal}, which is based on structural equations. In this work, we focus on Rubin's approach. The characteristic of Rubin's approach is the notion of the potential outcome. Suppose we investigate the effect of medicine X on patient A. We consider (i) the health status (e.g., blood pressure $x_T \in \mathbb{R}$) of patient A if he/she took medicine X and (ii) the health status $x_C \in \mathbb{R}$ of patient A if he/she did not take medicine X. Rubin's model defines the causal effect of medicine X as the difference of the two results, i.e., $x_T - x_C$. More formally, in this framework, we have a population $\mathcal{X}$ and intervene each individual $x \in \mathcal{X}$. The way of intervention is selected from a predefined set, e.g., prescribing medicine A or B or C. The interventions are called treatments in this framework. The result is measured by a real number which is called an outcome. The causal effect is measured as the difference in the outcome values.

In the citation-aware publication venue recommendation problem, an individual corresponds to a paper, a treatment corresponds to publishing the paper in a venue, and the outcome corresponds to the number of citations. The aim is to estimate the causal effect of venue selection.

The challenge in this framework is that we cannot observe both of $x_T$ and $x_C$, i.e., the fundamental problem of causal inference. Many methods have been proposed to tackle this problem. The meta-strategy \citep{kunzel2019metalearners} employed in this work is one of them.

Following this framework, we assume that samples $(\boldx, t, \{y(s)\}_{s \in \mathcal{T}})$ are generated from an unknown distribution in an i.i.d. manner, where $\boldx$ is the feature of the paper, $t$ is the venue where the paper is published, and $y(s)$ is the number of citations that the paper will receive when the paper is published in venue $s \in \mathcal{T}$. In this framework, we observe only $y(t)$, and we cannot observe $\{y(s)\}_{s \neq t}$. The observed outcome $y(t)$ is called the factual outcome, and the unobserved outcomes $\{y(s)\}_{s \neq t}$ are called counterfactual outcomes. Although it would be an ideal to recommend $\argmax_{s \in \mathcal{T}} y(s)$, $y(s)$ is not completely observable even in the training time. Therefore, it is impossible to directly learn the best recommendation function. We solve this issue by employing the treatment effect estimation framework in the following sections.

\begin{table}[tb]
    \centering
    \caption{Notations.}
    \begin{tabular}{ll} \toprule
        Notations                              & Descriptions \\ \midrule
        $\boldx_i \in \mathbb{R}^d$            & The feature vector of paper $i$. \\
        $\boldX$                               & The random variable of feature vectors. \\
        $\mathcal{T}$                          & The set of publication venues. \\
        $t_i \in \mathcal{T}$                  & The publication venue of paper $i$. \\
        $T$                                    & The random variable of publication venues. \\
        $y_i \in \mathbb{R}_+$                 & The number of citations paper $i$ have received. \\
        $Y(s)$                                 & The random variable of the number of citations a paper receives in venue $s \in \mathcal{T}$. \\
        $Y^F$                                  & The random variable of the number of citations a paper receives in the factual venue. \\
        $r\colon \mathbb{R}^d \to \mathcal{T}$ & A publication venue recommender system. \\
        $\mathcal{D}$                          & The training data. \\
        $\mu_t(\boldx)$                        & The control response function $\mu_t(\boldx) = \mathbb{E}[Y(t) \mid \boldx]$. \\
        \bottomrule
    \end{tabular}
    \label{tab: notations}
\end{table}

Let $\boldX$ denote the random variable of paper features, $T$ denote the random variable of treatments (i.e., venue), $Y(t)$ denote the random variable of the potential outcome when the paper is published in venue $t \in \mathcal{T}$, and $Y^F$ denote the random variable of the factual outcome. Notations are summarized in Table \ref{tab: notations}.

We assume that the contribution of the paper receiving recommendations is sufficiently significant and that the reviewing process is so reasonable that the paper is accepted if it is submitted to an appropriate venue. Admittedly, this assumption does not always hold in reality, and a paper can be rejected by a publication venue even if the paper has significant potential influence in the venue. When a user of the system feels that the research at hand is not significant enough, he/she can manually filter out challenging venues and input only reasonable (not-so-challenging) venues to the system. Then, the recommender system would suggest the best venue among the reasonable venues. This mitigates the above-mentioned rejection problem. Alternatively, we can build another model that predict the probability that a paper is accepted to each venue, and estimate the score of the venue based on both the acceptance probability and potential impact. This will reduce the trouble of manual filtering process. In this work, we focus on modeling potential impact of venue selection. Considering acceptance probability for recommending where to \emph{submit} instead of where to \emph{publish} in an end-to-end manner is an important next step of this work.

Besides, there are many more factors when we decide the review, such as smoothness, speed, and quality of the reviewing process, and the citation count is by no means the only factor for choosing publication venues. We do not expect \textsc{Poincare} is in full charge of venue decision. Rather, \textsc{Poincare} tells one aspect of the decision process. To take other factors into consideration, we can choose the initial candidate venues based on these important factors. For example, we can use an existing association-based recommender system or manually select venues in this phase. After we list comparative venue candidates, we can consult \textsc{Poincare} in which venue among the selected candidates the research attracts the most attention and decide the final venue based (partially) on the impact-aware recommendation. 

The important fact is that \textsc{Poincare} can provide novel (i.e., impact-aware) aspects that traditional recommender systems cannot provide. We can also use \textsc{Poincare} along with association-based recommender systems, i.e., simultaneously consult both \textsc{Poincare} and off-the-shelf association-based recommender systems and decide the venue based on both aspects. The use of \textsc{Poincare} broadens the perspective in the venue decision process compared to solely relying on association-based recommender systems.

\section{Related Work}

\subsection{Treatment Effect Estimation}
The estimation of treatment effects from observational data is a critical research topic in science \citep{spirtes2000causation, imbens2015causal, morgan2015counterfactuals, pearl2009causality}.
In this study, we employ the Rubin causal model \citep{rubin1974estimating}, where we consider potential outcomes. Notice that it is impossible to observe counterfactual potential outcomes. For example, in the case of citation prediction, we can observe the number of citations in a factual venue but cannot observe the number of citations in the other (counterfactual) venues. Causal inference frameworks have been employed in the informetrics literature. For example, Traag \citep{traag2021inferring} employed Bayesian modeling and a causal inference perspective to investigate the influence of journals. Davis et al. \citep{davis2008open} conducted randomized controlled trials to estimate the effect of open access publishing. To the best of knowledge, this paper is the first to recommend publication venues using the treatment effect estimation framework.
The treatment effect estimation has many applications in the machine learning and data mining community including advertisement \citep{dalessandro2012causally, perlich2012bid, wang2014efficient, sun2015causal, wang2015robust, li2016matching}, education \citep{zhao2017estimating, olaya2020uplift}, and recommendation \citep{schnabel2016recommendations, sharma2015estimating, bonner2018causal, gilotte2019offline, joachims2016counterfactual}. For example, in the advertisement setting, we want to estimate the sales of a product with and without an advertisement and to decide whether we should place an advertisement. There are plenty of machine learning and statistical methods to estimate treatment effects. For example, Johansson et al. \citep{johansson2016learning} and Shalit et al. \citep{shalit2017estimating} proposed to learn effective representations that have similar distributions between the treatment and control population. Schnabel et al. \citep{yoon2018ganite} proposed to use generative adversarial networks \citep{goodfellow2014generative} to estimate treatment effects. Schnabel et al. \citep{schnabel2016recommendations} correct biases by the empirical risk minimization of the inverse propensity scoring. K{\"u}nzel et al. \citep{kunzel2019metalearners} proposed some strategies (meta-learners) for estimating treatment effects. Although most of the existing works focused on the binary treatment setting, several previous works tackled multiple treatment settings \citep{rzepakowski2012decision, zhao2017uplift, zhao2017practically} as in the setting of this paper. Pearl \citep{pearl2009causal} provided a survey for causal inference, and Yao et al.'s survey \citep{yao2020survey} was from the perspective of machine learning and data mining. In this paper, we develop an effective academic venue recommender system using an off-the-shelf framework develop in the machine learning literature.

\subsection{Publication Venue Recommendation}
Recently, several recommender systems have been proposed for academic publication venues. For example, Yang and Davison \citep{yang2012venue} recommend venues where similar papers are published. They utilize collaborative filtering for recommending venues. Yang et al. \citep{yang2014recommendation} improve the recommendation performance by solving venue, paper, and co-author recommendations simultaneously. The main benefit of these methods is their effectiveness owing to the rich data they used. By contrast, we use only fields of study data. The main advantage of our approach is that our data are easy to obtain and maintain, which is a critical benefit when we deploy and maintain the recommender system. However, it would be an important future direction to boost the effectiveness of our method using rich data employed in \citep{yang2012venue, yang2014recommendation}. Medvet et al. \citep{medvet2014publication} extract topics from the title and abstract and recommend venues where papers with similar topics are published. In other words, their approach is content-based. Our work can be situated in this line of work as \textsc{Poincare} also employs a content-based approach. The differences between Medvet et al. and our method are three-fold. First, we directly use the data on the fields of study, while Medvet et al. estimated latent fields. The advantage of our approach is that it is more effective and easier to implement, while the advantage of the approach employed by Medvet et al. is its broad applicability, i.e., it does not require field data. Second, Medvet et al. employed a distance-based approach, while \textsc{Poincare} employed a statistic model. The advantage of our approach is its flexibility. We can combine \textsc{Poincare} with any other statistical machine learning models, such as support vector machines and neural networks. Last but not least, our approach directly handles selection biases of venue selection, while any existing methods do not. Chen et al. \citep{chen2015aver} and other authors \citep{luong2012publication, luong2012exploiting, alhoori2017recommendation, yu2018pave} recommend venues based on academic social networks. For example, Chen et al. \citep{chen2015aver} employ random walks on academic social networks and extract associations of venues and authors. Therefore, these systems tend to recommend venues where similar authors published papers. Feng et al. \citep{feng2019deep} used word embedding and deep neural networks to recommend publication venues in the biomedical field. Overall, the existing methods recommend the most likely venues, and they do not aim to maximize the influence of the paper. In other words, they are all association-based recommendation approaches. In contrast, we propose a potential influence-based recommendation method in this study.

\subsection{Citation Prediction}
Our proposed method can be formalized as a citation prediction method. The citation prediction problem has been studied in the informetrics community \citep{tahamtan2018core} and data mining community for a long time. For example, Yan et al. \citep{yan2011citation} predicted the number of citations in a certain period based on the features and contents of a paper. Note that this kind of citation prediction methods can be directly used for venue recommendation owing to our formulation of \emph{citation-aware publication venue recommendation problem} presented in Section 2. We assume that one important contribution of our work is that we bridge the citation prediction literature and venue recommender system literature via this formulation, and thereby enable us to exploit the rich existing results of citation prediction for improving venue recommendations. Yu et al. \citep{yu2012citation} proposed a meta-path based method to predict citations as a link prediction problem. Wang et al. \citep{wang2013quantifying} modeled long-term citation dynamics based on preferential attachment, aging, and fitness. Davletov et al. \citep{davletov2014high} predicted the number of citations in the future using not only temporal features but also topological features such as the betweenness centrality. Onodera et al. \citep{onodera2015factors} estimated the number of citations using various factors and found that the price index affects much. Abrishami et al. \citep{abrishami2019predicting} used deep encoder-decoder networks to predict the number of citations effectively. Shen et al. \citep{shen2014modeling}, Xiao et al. \citep{xiao2016modeling}, and Bai et al. \citep{bai2019predicting} modeled popularity dynamics based on a point process. Abramo et al. \citep{abramo2019predicting} predicted long-term impact based on early citations and journal impact factor. Dong et al. \citep{dong2015will} predicted whether a paper would increase the h-index of the authors. The philosophy of Dong et al. is common with our work, i.e., ``\emph{provide concrete suggestions to researchers for better expanding their scientific influence}'' (cited from \citep{dong2015will}.) Our approach can be seen as the next step of Dong et al. as they formulated the problem by a prediction problem, while we provide a concrete action via venue recommendation. The relationship between the citation pattern and the content of a paper has also been extensively studied \citep{vieira2010citations, falagas2013impact, buter2011non, subotic2014short}. Overall, the existing methods forecast the number of citations or model the transition of the number of citations, and they do not recommend venues to maximize the impact. In particular, they do not employ the treatment effect estimation framework, which is the most significant difference between our method and the existing citation prediction methods.

\section{Proposed Method} \label{sec: method}

In a nutshell, our proposed method, \textsc{Poincare}, considers choosing a publication venue as a \emph{treatment} and the number of citations a paper receives as an \emph{outcome}, and it estimates a potential outcome with the aid of a treatment effect estimation framework.
To build effective methods, we need to consider the following facts.

\begin{itemize}
    \item \textbf{Different laws.} Different venues have different preferences. For example, scalability is important in data mining and information retrieval conferences, while sample efficiency is important in machine learning conferences. Different preferences are confirmed in the experiments (Observations 4 and 9).
    \item \textbf{Selection biases.} There exist selection biases in choosing publication venues. This is confirmed in the experiments (Observation 1). 
\end{itemize}

We employ meta strategies of K{\"u}nzel et al. \citep{kunzel2019metalearners}, who proposed three strategies (meta-learners) for estimating treatment effects in the binary treatment setting. The T-learner models the control response function $\mu_t(\boldx) = \mathbb{E}[Y(t) \mid \boldx]$ for each $t = 0, 1$ using two models. The models that estimate the control response functions are called base learners. The S-learner includes the treatment indicator as a feature and models the response function $\mu(\boldx, t) = \mathbb{E}[Y(t) \mid \boldx]$ using a single base learner. The X-learner models the control response function separately as the T-learner, estimates individual treatment effects by plugging the estimated outcomes in the formula, and learns the estimated individual treatment effects. In the venue recommendation problem, we have more than two treatments (i.e., venues). Thus, we extend their meta strategies to multi-treatment settings, except for the X-learner, which is not immediately applicable to multi-treatment settings. Specifically, in multi-treatment settings, the T-learner has $|\mathcal{T}|$ base learners, one for each venue, and they estimate the control response functions $\mu_t(\boldx)$ for venues $t \in \mathcal{T}$. T-learner approximates the control response function by $\mu_t(\boldx) = \mathbb{E}[Y(t) \mid \boldx] \approx \mathbb{E}[Y^F \mid \boldx, t]$. This approximation is exact when the ignorability assumption holds \citep{kunzel2019metalearners}. For each treatment $t \in \mathcal{T}$, we can fit the base learners $\hat{\mu}_t$ using only factual outcomes $\{(\boldx_i, y_i) \mid t_i = t\}$ thanks to the approximation. The S-learner has a single base learner in the multi-treatment setting, and it models the response function $\mu(\boldx, t)$ as in the binary treatment setting.  Among the T- and S-learners, we employ the T-learner, which utilizes different response functions for different treatments, owing to the different laws mentioned above. We confirm that the T-learner is more effective than the S-learner in the experiments (Observation 4).

Any choices for base learners $\hat{\mu}_t$ can be combined with \textsc{Poincare}, from linear models to deep learning models. We adopt linear models for the base learners to ensure the interpretability of recommendations. 
After training base learners $\hat{\mu}_t$, \textsc{Poincare} recommends the venue that has the largest potential outcome, i.e., $r(\boldx) = \argmax_t \hat{\mu}_t(\boldx)$.

\begin{figure*}[t]
  \centering
\includegraphics[width=\hsize]{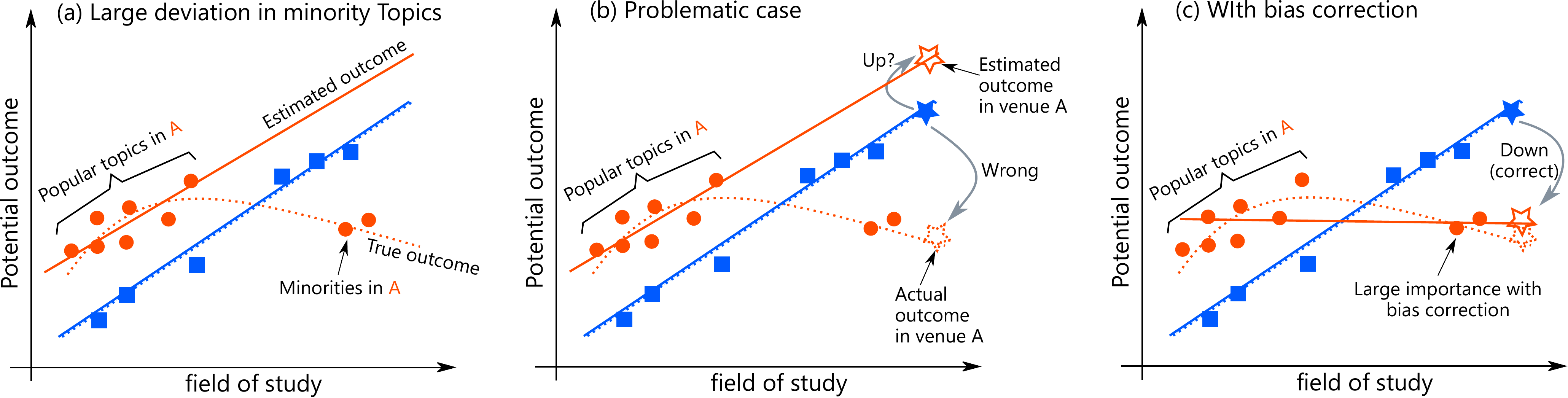}
\caption{(a) Large deviation: Each point represents a paper. Suppose infinitely many papers have been published in the popular topics in venue A. Without bias correction, estimated outcomes (represented by a solid red line) are fitted well to popular topics, but not to minority topics. (B) Problematic case: This deviation is problematic. Let us investigate the paper represented by a blue filled star published in venue B. The estimator tells that this paper would have received more citations in venue A. However, this estimation deviates from the truth, and actually, it would have received fewer citations in venue A. Similar arguments follow for similar papers in venue B. (c) With bias correction: If we correct the bias, the estimator puts importance on the minority papers as well and recommends venue B for the star paper correctly.}
 \label{fig: illust2}
\end{figure*}

As mentioned above, there exist selection biases in choosing publication venues. Without correcting the biases, the base learners may fit only to a majority of papers in the venue and ignore minority papers in the venue. However, the aim of treatment effect estimation is to estimate the number of citations in \emph{counterfactual} venues. Figure \ref{fig: illust2} illustrates why bias correction is required. We employ bias correction methods to estimate counterfactual outcomes accurately.
Namely, we utilize the inverse propensity weighting loss \citep{schnabel2016recommendations}. The propensity score $\text{P}[t \mid \boldx]$ is the probability that covariate $\boldx$ receives treatment $t$. The inverse propensity weighting loss puts more importance on unlikely papers, i.e., the papers with small propensity scores. This is intuitively because small propensity scores indicate that many similar papers are published in other conferences, and small propensity papers are important for \emph{counterfactual} treatment effect estimations of the similar papers.
Formally, the loss function is written as

\begin{align} \label{eq: loss_est}
    \mathcal{L}_{\text{est}} = \frac{1}{|\mathcal{D}|} \sum_{(\boldx_i, t_i, y_i) \in \mathcal{D}} \frac{1}{\hat{\text{P}}[t_i \mid \boldx_i]} (y_i - \hat{\mu}_{t_i}(\boldx_i))^2,
\end{align}

\noindent where $\hat{\text{P}}[t_i \mid \boldx_i]$ is the estimated propensity score. We estimate the propensity score using logistic regression following \citep{rosenbaum1983central, schnabel2016recommendations}. Specifically, the logistic regression model $\hat{\text{P}}$ is trained on the training dataset $\{(\boldx_i, t_i) \mid (\boldx_i, t_i, y_i) \in \mathcal{D}\}$, where the covariates $\boldx_i$ are the inputs and the treatment $t_i$ is the class label, constructed from $\mathcal{D}$.
In addition to the regression error, we regularize the parameters to prevent overfitting. Therefore, the final objective function is

\begin{align} \label{eq: loss}
    \mathcal{L} = \mathcal{L}_{\text{est}} + \lambda \|\boldtheta\|_2^2,
\end{align}

\noindent where $\lambda$ is a hyperparameter and $\boldtheta$ denotes the parameters of the base learner. It should be noted that when the base learners are linear models, Eq.~\ref{eq: loss} can be optimized using off-the-shelf implementations for ridge regression, such as scikit-learn \citep{pedregosa2011scikit}, by setting the sample weight equal to $1 / \hat{\text{P}}[t_i \mid \boldx_i]$. The algorithm is summarized in Algorithm \ref{algo}.

\begin{algorithm2e}[t]
\caption{\textsc{Poincare}}
\label{algo}
\DontPrintSemicolon 
\nl\KwData{Training dataset $\mathcal{D} = \{(\boldx_i, t_i, y_i)\}_{1 \le i \le n}$, Queries $\mathcal{Q} = \{\boldq_j\}_{1 \le j \le m}$.}
\nl\KwResult{Recommended venues $\{t_j\}_{1 \le j \le m}$.}
    \nl Train the propensity score model $\hat{\text{P}}$ by $\{(\boldx_i, t_i) \mid (\boldx_i, t_i, y_i) \in \mathcal{D}\}$. \;
    \nl Estimate sample weights in Eq.~\ref{eq: loss_est} by $\hat{\text{P}}$. \;
    \nl Train estimators $\hat{\mu}_t$ by Eq.~\ref{eq: loss}. \;
    \nl $t_j = \argmax_t \hat{\mu}_t(\boldq_j) \quad \forall j \in \{1, \cdots, m\}$. \;
\end{algorithm2e}

\section{Experiments}

We confirm the effectiveness of \textsc{Poincare} using a paper dataset from computer science conferences and illustrative simulated data.

\subsection{Dataset} \label{sec: dataset}

\begin{table}[t]
    \centering
    \caption{Dataset statistics.}
    \scalebox{0.8}{
    \begin{tabular}{crrrrrr} \toprule
        & AAAI & IJCAI & KDD & NeurIPS & ICML & Total \\ \midrule
        Number of papers & $634$ & $605$ & $172$ & $528$ & $368$ & $2307$ \\
        Average citations & $17.0$ & $14.5$ & $37.0$ & $66.1$ & $84.7$ & $39.9$ \\
        Field & Artifical Intelligence & Artificial Intelligence & Data Mining & Machine Learning & Machine Learning &  \\ \bottomrule
    \end{tabular}
    }
    \label{tab: dataset}
\end{table}

We construct a dataset for the citation-aware publication venue recommendation problem based on the dblp v12 dataset\footnote{\url{https://www.aminer.org/citation}} \citep{tang2008arnetminer}. We use five computer science conferences, $\mathcal{T} = \{$AAAI, IJCAI, KDD, NeurIPS, ICML$\}$, and extract accepted papers in 2015 from the dblp dataset. Let $\mathcal{F}$ denote the set of fields of study in the dblp dataset, such as ``Recommender system,'' ``Causal inference,'' and ``Debiasing.''
We construct the feature vector $\boldx_i \in \mathbb{R}^\mathcal{F}$ of paper $i$ based on the fields of paper $i$. The $f$-th dimension $\boldx_{i, f}$ is $1$ if $f \in \mathcal{F}$ is a field of study of paper $i$, and $0$ otherwise. We split the dataset into training and test datasets in a stratified fashion. To be specific, for each venue, we use random $70$ percents of papers in the training dataset and the others in the test dataset. Table \ref{tab: dataset} summarizes the statistics. The preprocessed data are available at \url{https://github.com/joisino/poincare}.

\subsection{Selection Biases} \label{sec: bias}

In this section, we investigate the existence of selection biases in choosing publication venues. To this end, we conduct two-sample tests between the covariate distributions $\text{P}[\boldx \mid t]$ of venues $t$. Specifically, we utilize the kernel two-sample test \citep{gretton2012kernel}. For each venue $t \in \mathcal{T}$, let $\mathcal{D}_t = \{\boldx_i \mid (\boldx_i, t_i, y_i) \in \mathcal{D}, t_i = t\}$ be the set of papers published in venue $t$. For each pair $\{s, t\} \subseteq \mathcal{T}$ of venues, we first compute the unbiased estimator of the maximum mean discrepancy (MMD) \citep{gretton2012kernel}, defined as

\begin{align} \label{eq: MMD}
    \widehat{\text{MMD}}^2 &= \frac{1}{|\mathcal{D}_s|(|\mathcal{D}_s| - 1)} \sum_{\bolds_i, \bolds_j \in \mathcal{D}_s, i \neq j} k(\bolds_i, \bolds_j) \\
    &+ \frac{1}{|\mathcal{D}_t|(|\mathcal{D}_t| - 1)} \sum_{\boldt_i, \boldt_j \in \mathcal{D}_t, i \neq j} k(\boldt_i, \boldt_j) \notag \\
    &- \frac{2}{|\mathcal{D}_s| |\mathcal{D}_t|} \sum_{\boldt_i \in \mathcal{D}_s, \boldt_j \in \mathcal{D}_t} k(\bolds_i, \boldt_j), \notag
\end{align}

\noindent where $k$ is a kernel function. We use the Gaussian kernel $k(\boldx, \boldx') = \exp(-\|\boldx - \boldx' \|_2^2 / (2 \sigma^2))$ and set the bandwidth $\sigma$ to the median of distances. Table \ref{tab: MMD} reports the unbiased estimators of MMDs between venues. We can observe that AAAI and IJCAI, NeurIPS and ICML are similar in terms of covariate distributions. This is reasonable because AAAI and IJCAI are both artificial intelligence conferences, and NeurIPS and ICML are both machine learning conferences. However, even these pairs have high MMD values. We validate this quantitatively via statistical tests.

\begin{table}[t]
    \centering
    \caption{Selection biases: MMD ($\times 10^3$) between venues. $^*$ denotes statistical significance with significance level $\alpha = 0.01$. This result shows that different venues have different tendencies of publications.}
    \begin{tabular}{crrrrr} \toprule
        & AAAI & IJCAI & KDD & NeurIPS & ICML \\ \midrule
        AAAI & - & - & - & - & -\\
        IJCAI & $9.8^*$ & - & - & - & -\\ 
        KDD & $20.7^*$ & $33.3^*$ & - & - & - \\ 
        NeurIPS & $10.7^*$ & $19.6^*$ & $33.7^*$ & - & - \\ 
        ICML & $10.4^*$ & $20.0^*$ & $38.5^*$ & $4.0^*$ & - \\ 
        \bottomrule
    \end{tabular}
    \label{tab: MMD}
\end{table}

The exact MMD is zero if and only if two distributions $P$ and $Q$ are identical, and the asymptotic distribution of the unbiased estimation of MMD (Eq.~\ref{eq: MMD}) follows the chi-square like distribution under the null hypothesis $P = Q$, as shown in \citep{gretton2012kernel, gretton2009fast}. We estimate the distribution under the null hypothesis by sampling, and we conduct statistical tests. Table \ref{tab: MMD} indicates that all null hypothesizes are rejected. This indicates that covariate distributions are indeed different from venue to venue, even between NeurIPS and ICML. In other words, selection biases exist.

\begin{description}
    \item[Observation 1.] There exist selection biases in choosing publication venues.
\end{description}

\subsection{Association Does Not Indicate Impact} \label{sec: corr}

Association-based venue recommender systems recommend venues where a paper is likely to be published. We confirm that a paper does not necessarily receive many citations even if that paper is very likely to be published in that venue.

\subsubsection{Experimental setup}
In the experiments, the input of \textsc{Poincare} is the feature vector $\boldx_i$, and the target is the logarithm $\log(y_i)$ of the number of citations. We select the regularization coefficient $\lambda \in [0.001, 90]$ in Eq.~\ref{eq: loss} by $5$-fold cross validation, and we train the model with the entire training dataset using the chosen hyperparameters in the test time.

\subsubsection{Variants} We use two variants of \textsc{Poincare} to validate the choice of building blocks of \textsc{Poincare}. The first one is \textsc{Poincare}-UW (uniform weight), which does not use the inverse propensity loss. This corresponds to \textsc{Poincare} with $\hat{\text{P}}$ being a constant in Eq.~\ref{eq: loss_est}. The second one is \textsc{Poincare}-S (single), which employs the S-learner as the meta-learner instead of the T-learner. This corresponds to \textsc{Poincare} that shares one model for all $\hat{\mu}_t$ and takes venue $t$ as an input feature. The hyperparameter for these variants are also selected by $5$-fold cross validation.

\subsubsection{Baselines} We use four association-based baseline models, namely, linear regression model, random forest model, support vector machine, and multilayer perceptron. These models are trained using a set of pairs of a feature vector and venue $\{(\boldx_i, t_i) \in \mathbb{R}^{d} \times \mathcal{T}\}_{i = 1, \dots, n}$, and they recommend the venue the likelihood of which is the highest. The hyperparameters for baselines are selected by $5$-fold cross validation. The ranges of hyperparameters are the following, \texttt{C} $\in [0.001, 90]$ for logistic regression, \texttt{n\_estimators} $\in \{10, 100, 1000\}$, \texttt{max\_depth} $\in \{2, 3, \texttt{None}\}$, \texttt{criterion} $\in \{\text{gini}, \text{entropy}\}$, \texttt{min\_samples\_split} $\in \{2, 3, 5\}$, \texttt{min\_samples\_leaf} $\in \{1, 2, 5\}$ for random forest, \texttt{C} $\in [0.001, 90]$, \texttt{kernel} $\in \{\text{poly}, \text{rbf}, \text{sigmoid}\}$ for support vector machine, \texttt{hidden\_layer\_sizes} $\in \{(32,), (64,), (128,), (256,)\}$, \texttt{alpha} $\in \{0.01, 0.001, 0.0001, 0.00001\}$, \texttt{learning\_rate\_init} $\in \{0.001, 0.0001, 0.00001\}$ for multi layer perceptron. These notations and variables follow the scikit learn package. We used the \texttt{GridSearchCV} procedure of the scikit-learn package to choose hyperparameters. We set other hyperparameters to the default values. Note that existing methods, such as AVER \citep{chen2015aver} and PAVE \citep{yu2018pave}, cannot be used here directly because they use different information, such as abstract texts and author information, compared to the fields of studies in this study. Nevertheless, the consistent results in this experiment indicate that these results can be extended to other settings.

\begin{figure}
 \begin{minipage}{0.48\hsize}
  \centering
    \includegraphics[width=\hsize]{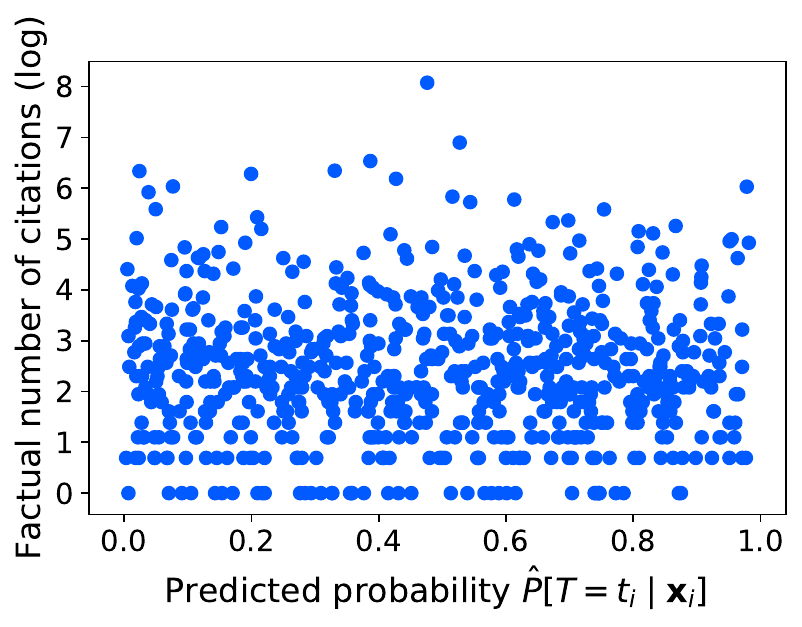}
 \end{minipage}
 \hfill
 \begin{minipage}{0.48\hsize}
  \centering
    \includegraphics[width=\hsize]{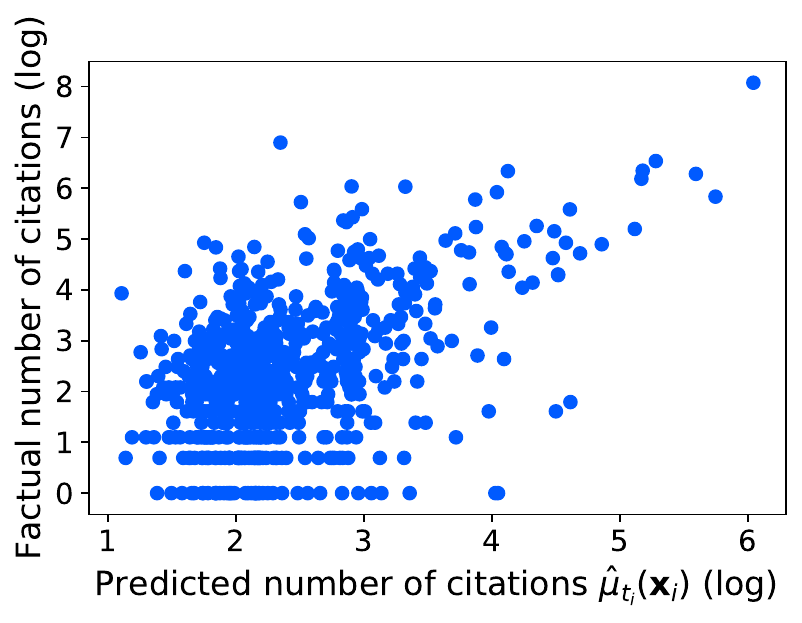}
 \end{minipage}
    \caption{X-axis: The scores of the factual conferences given by (Left) the linear regression association-based model and (Right) \textsc{Poincare}. Y-axis: The factual numbers of citations that papers received. The scores of \textsc{Poincare} are more correlated with the number of citations (i.e., impact) than those of the association-based model. }
 \label{fig: corr}
\end{figure}

\subsubsection{Evaluation} Both \textsc{Poincare} and the association-based methods estimate a score $f(t)$ for each venue $t \in \mathcal{T}$ and recommend the venue $\argmax_{t \in \mathcal{T}} f(t)$ with the highest score. The score of \textsc{Poincare} is the predicted number of citations the paper receives when the paper is published in the venue. The score of association-based methods is the predicted probability $\text{P}[T = t \mid \boldx]$ that the paper is published in the venue. 
For each method, we compute the Spearman's rank correlation coefficient between the scores and the numbers of citations in the factual venue using the papers in the test dataset. High correlation coefficients indicate that the score reflects the influence of the paper.

\begin{table*}[tb]
    \centering
    \caption{Correlation with impact: The Spearman's correlation coefficient between the predicted scores and numbers of citations. The first four models are association-based models, and the following three models are variants of the proposed method. \newline $^*$ denotes statistically significant correlation with the t-test with significance level $\alpha = 0.01$. The highest score and scores within $\pm$ stddev of the highest score are marked in \textbf{bold}.}
    \scalebox{0.8}{
    \begin{tabular}{ccccccc} \toprule
        & \multicolumn{1}{c}{AAAI} & \multicolumn{1}{c}{IJCAI} & \multicolumn{1}{c}{KDD} & \multicolumn{1}{c}{NeurIPS} & \multicolumn{1}{c}{ICML} & \multicolumn{1}{c}{Total} \\ \midrule
        Logistic regression    & 0.000 $\pm$ 0.058       & -0.081 $\pm$ 0.046      & \bft{0.109 $\pm$ 0.097} & 0.117 $\pm$ 0.060       & -0.030 $\pm$ 0.066      & -0.028 $\pm$ 0.026 \\
        Random forest          & 0.015 $\pm$ 0.064       & -0.096 $\pm$ 0.045      & 0.069 $\pm$ 0.150       & 0.120 $\pm$ 0.084       & 0.048 $\pm$ 0.059       & -0.094 $\pm$ 0.029 \\
        Support vector machine & 0.004 $\pm$ 0.070       & -0.081 $\pm$ 0.050      & 0.073 $\pm$ 0.105       & 0.160 $\pm$ 0.049       & 0.058 $\pm$ 0.074       & 0.025 $\pm$ 0.031 \\
        Multilayer perceptron  & 0.006 $\pm$ 0.046       & -0.063 $\pm$ 0.049      & 0.087 $\pm$ 0.107       & 0.143 $\pm$ 0.040       & 0.017 $\pm$ 0.068       & 0.014 $\pm$ 0.024 \\ \midrule
        \textsc{Poincare}      & \bft{0.179 $\pm$ 0.050} & \bft{0.252 $\pm$ 0.078} & \bft{0.159 $\pm$ 0.085} & \bft{0.406 $\pm$ 0.041} & \bft{0.457 $\pm$ 0.071} & \bft{0.388 $\pm$ 0.026}$^*$ \\
        \textsc{Poincare}-UW   & \bft{0.183 $\pm$ 0.049} & \bft{0.245 $\pm$ 0.071} & \bft{0.164 $\pm$ 0.073} & \bft{0.405 $\pm$ 0.040} & \bft{0.460 $\pm$ 0.065} & \bft{0.389 $\pm$ 0.024}$^*$ \\
        \textsc{Poincare}-S    & 0.033 $\pm$ 0.068       & 0.076 $\pm$ 0.077       & \bft{0.110 $\pm$ 0.108} & 0.343 $\pm$ 0.063       & \bft{0.457 $\pm$ 0.071} & 0.192 $\pm$ 0.035$^*$ \\
        \bottomrule
    \end{tabular}
    }
    \label{tab: corr}
\end{table*}

\subsubsection{Results}  Table \ref{tab: corr} reports the average and standard deviation of the correlation coefficients for $10$ different random seeds. This indicates that the scores of \textsc{Poincare} correlate with the numbers of citations, while the association-based methods do not correlate with the numbers of citations. In particular, all the association-based methods do not reject the null hypothesis with significance level $\alpha = 0.01$, while \textsc{Poincare} statistically significantly correlates with the impact ($p < 10^{-10}$). Figure \ref{fig: corr} plots the scores and numbers of citations of the logistic regression model and \textsc{Poincare}. They show that many papers receive few citations even if they are published in likely venues, and many other papers receive many citations even if they are published in unlikely venues, while \textsc{Poincare} can predict the number of citations effectively. We would like to remark that it does not necessarily mean that \textsc{Poincare} is always better than association-based methods, but it just indicates that association-based methods are not suitable for the citation-aware publication venue recommendation problem. If we want to recommend papers to likely venues, then association-based methods are suitable, while if we want to recommend paper to maximize the impact, as in our settings, \textsc{Poincare} is beneficial. As for the variants of \textsc{Poincare}, we observe that \textsc{Poincare} performs better than \textsc{Poincare}-S in Table \ref{tab: corr}. This result validates the choice of the T-learner. The performances of \textsc{Poincare} and \textsc{Poincare}-UW reported in Table \ref{tab: corr} are almost the same. This is because the observational data we use for evaluation are also biased, as seen in Section \ref{sec: bias}.
We need counterfactual evaluation to distinguish effectiveness of \textsc{Poincare} and \textsc{Poincare}-UW.
Unfortunately, it is impossible to evaluate methods in unbiased and counterfactual settings because we cannot observe counterfactual outcomes. In the next section, we illustrate the effectiveness of \textsc{Poincare} over \textsc{Poincare}-UW in counterfactual settings using synthetic datasets.

\begin{description}
    \item[Observation 2.] High likelihood $\text{P}[T = t_i \mid \boldx_i]$ does not necessarily indicate many citations.
    \item[Observation 3.] \textsc{Poincare} and \textsc{Poincare}-UW perform well in citation prediction.
    \item[Observation 4.] The T-learner based model (i.e., \textsc{Poincare}) performs better than the S-learner based model (i.e., \textsc{Poincare}-S).
\end{description}

\subsection{Simulation} \label{sec: syn}

\begin{table}[t]
    \centering
    \caption{Synthetic datasets. The highest score and scores within $\pm$ stddev of the highest score are marked in \textbf{bold}. $^*$ denotes statistically significant improvement over other methods with the paired t-test with significance level $\alpha = 0.01$. In particular, \textsc{Poincare} outperforms \textsc{Poincare}-UW in the counterfactual setting. }
    \begin{tabular}{ccc} \toprule
        & \multicolumn{1}{c}{Accuracy} & \multicolumn{1}{c}{Outcome ($\times 10^{-3}$)} \\ \midrule
        Logistic regression    & 0.7172 $\pm$ 0.0085 & 16.74 $\pm$ 0.65 \\
        Random forest          & 0.7092 $\pm$ 0.0082 & 16.57 $\pm$ 0.66 \\
        Support vector machine & 0.7163 $\pm$ 0.0075 & 16.71 $\pm$ 0.63  \\
        Multilayer perceptron & 0.7158 $\pm$ 0.0086 & 16.68 $\pm$ 0.63  \\ \midrule
        \textsc{Poincare}      & \bft{0.9385 $\pm$ 0.0059}$^*$ & \bft{20.60 $\pm$ 0.48}$^*$ \\
        \textsc{Poincare}-UW   & 0.8592 $\pm$ 0.0040 & 19.76 $\pm$ 0.49 \\
        \textsc{Poincare}-S    & 0.4996 $\pm$ 0.0059 & 10.44 $\pm$ 0.48 \\
        \bottomrule
    \end{tabular}
    \label{tab: syn}
\end{table}

Although the results in the previous section indicate the promising effectiveness of \textsc{Poincare} over association-based recommender systems, they could not distinguish the effectiveness of \textsc{Poincare} and \textsc{Poincare}-UW due to the factual evaluation. In this section, we generate a toy synthetic dataset to illustrate the effectiveness of \textsc{Poincare} over \textsc{Poincare}-UW. By synthesizing a dataset, we can know counterfactual outcomes and carry out counterfactual evaluations. Note that we do not use counterfactual outcomes when we train our models. We use them only for evaluation.

\subsubsection{Dataset synthesis} We synthesize a toy dataset with $10000$ ``papers'' and two ``venues.'' First, each paper is assigned to venue $t = -1$ with probability $0.5$ and venue $t = 1$ otherwise. The covariates $\boldx$ of the papers are drawn as $\boldx \sim \mathcal{N}(t \mathbbm{1}_d, 4 I_d)$, where $\mathbbm{1}_d$ is the vector of ones and $I_d$ is the identity matrix. In other words, papers with venue $t = -1$ tend to have negative values and papers with venue $t = 1$ tend to have positive values. This tendency reflects the selection biases of venues. We set the dimensions to $d = 16$. For each venue, we draw a random cross-term matrix $\boldA_t \sim \text{Unif}(0, 1)^{d \times d}$ and a linear-term vector $\boldb_t \sim \text{Unif}(0, 1)^d$, where each dimension is independent of other dimensions. We use different cross-term matrices and liner-term vectors for different venues to reflect the different laws of venues. Then, we compute the outcomes by $y(t) = \exp(0.01 \boldx^\top \boldA_{t} \boldx + \boldb_{t}^\top \boldx)$. The exponential function reflects various orders of citation counts. Note that the target of the base learners are the inside of the exponent because we take the logarithm in the preprocessing (see the experimental setup in Section \ref{sec: corr}).

\subsubsection{Experimental setup} We use the same settings and baselines as in Section \ref{sec: corr}.

\subsubsection{Evaluation} We evaluate methods by accuracy and average outcomes. Accuracy is defined as the proportion of papers that are recommended the venue with the highest ground truth outcome. The average outcome is the average value of the outcomes in the recommended venues. Higher values are better in both metrics.

\subsubsection{Results} Table \ref{tab: syn} shows the results. We can observe that \textsc{Poincare} and \textsc{Poincare}-UW perform better than association-based methods as indicated in Section \ref{sec: corr}. Besides, unlike in Section \ref{sec: corr}, \textsc{Poincare} outperforms \textsc{Poincare}-UW in this experiment. This is because we evaluate methods in counterfactual settings. These improvements are statistically significant with significant level $\alpha = 0.01$.
The results in this section illustrate the effectiveness of \textsc{Poincare} over \textsc{Poincare}-UW.

\begin{description}
    \item[Observation 5.] \textsc{Poincare} can perform better than \textsc{Poincare}-UW in counterfactual settings.
\end{description}

\subsection{Agreement with Influential Researchers} \label{sec: agreement}

In this section, we investigate the effectiveness of \textsc{Poincare} from a different perspective, i.e., whether the \textsc{Poincare}'s recommendations resemble decision making by influential researchers or by inexperienced researchers. 

\subsubsection{Experimental setup} First, we rank the last authors of the papers in the test dataset by the average number of citations their papers received. These values are computed with all papers they have ever published in the dblp dataset. We consider that a researcher is influential and experienced if he/she receives many citations on average. We use last authors because they tend to be principal investigators and advise and/or decide the publication venues of papers. We consider the top-half researchers are influential, and the bottom-half researchers are non-influential. We say that a recommender system agrees with the author of a paper if the recommendation is the same as the venue where the paper was actually published. For each recommender system, we gather the papers the recommender system agrees with, and investigate the ratio of influential researchers in these papers.

\begin{table*}[t]
    \centering
    \caption{Agreement with influential researchers: The ratio of papers written by influential researchers each recommender system agrees. The highest score and scores within $\pm$ stddev of the highest score are marked in \textbf{bold}. $^*$ denotes statistically significant improvement over other methods with the paired t-test with significance level $\alpha = 0.01$.}
    \begin{tabular}{ccccc} \toprule
        \multicolumn{1}{c}{Logistic regression} & \multicolumn{1}{c}{Random forest} & \multicolumn{1}{c}{Support vector machine} & \multicolumn{1}{c}{Multilayer perceptron} & \multicolumn{1}{c}{\textsc{Poincare}} \\ \midrule
        0.541 $\pm$ 0.018 & 0.520 $\pm$ 0.013 & 0.551 $\pm$ 0.015 & 0.553 $\pm$ 0.016 & \bft{0.753 $\pm$ 0.031}$^*$ \\
        \bottomrule
    \end{tabular}
    \label{tab: agree}
\end{table*}

\subsubsection{Results} Table \ref{tab: agree} reports the average and standard deviation of the agreement ratios for $10$ different random seeds. This indicates that \textsc{Poincare} agrees with more influential researchers than non-influential researchers. This ratio is statistically larger than association-based methods with a significance ratio $\alpha = 0.01$. This result also shows that the behavior of influential researchers is indeed different from that of non-influential researchers because \textsc{Poincare} distinguishes them and treats them differently. It indicates the room for reconsidering publication venues decided by researchers.
We stress that our ultimate goal is not to imitate influential researchers but to recommend suitable venues which might not be found by even influential researchers. Nevertheless, this experimental result highlights the effectiveness of our recommender system over association-based methods.

\begin{description}
    \item[Observation 6.] \textsc{Poincare} agrees with more influential researchers than non-influential researchers.
\end{description}

\subsection{Counterfactual Recommendations}

We investigate whether \textsc{Poincare} just follows factual venues or recommends counterfactual venues. We investigated the recommendations for $52$ papers published in KDD in the test dataset and found that $26$ papers were recommended other venues than KDD by \textsc{Poincare}. This indicates that \textsc{Poincare} agrees with half of KDD papers, but the other half of papers would have potential influence if they were published in other conferences according to \textsc{Poincare}.

\begin{description}
    \item[Observation 7.] \textsc{Poincare} recommends counterfactual venues as well.
\end{description}

\subsection{Interpretability}

\begin{figure*}[tb]
 \begin{minipage}{0.19\hsize}
  \centering
    \includegraphics[width=\hsize]{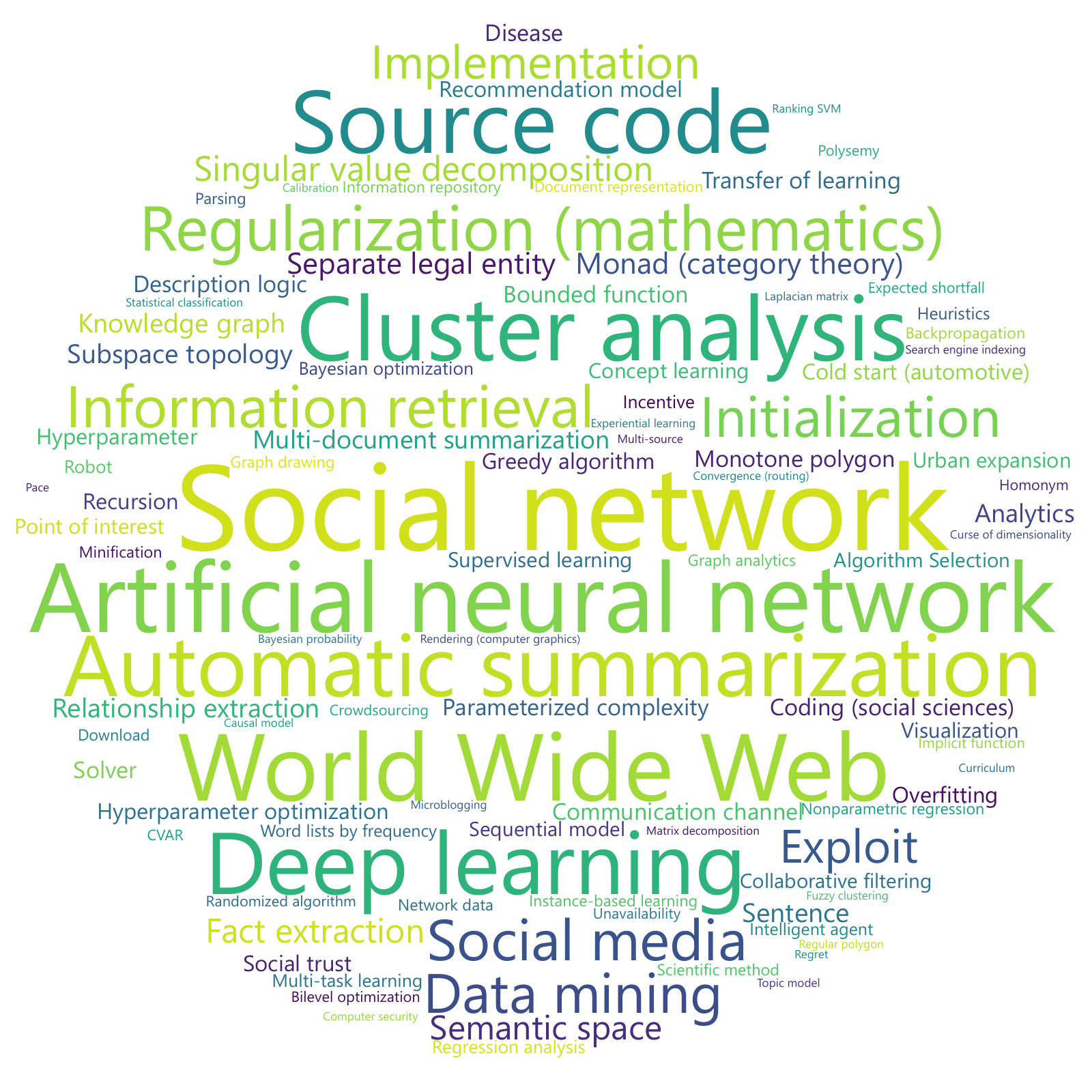}
    (a) AAAI
 \end{minipage}
 \begin{minipage}{0.19\hsize}
  \centering
    \includegraphics[width=\hsize]{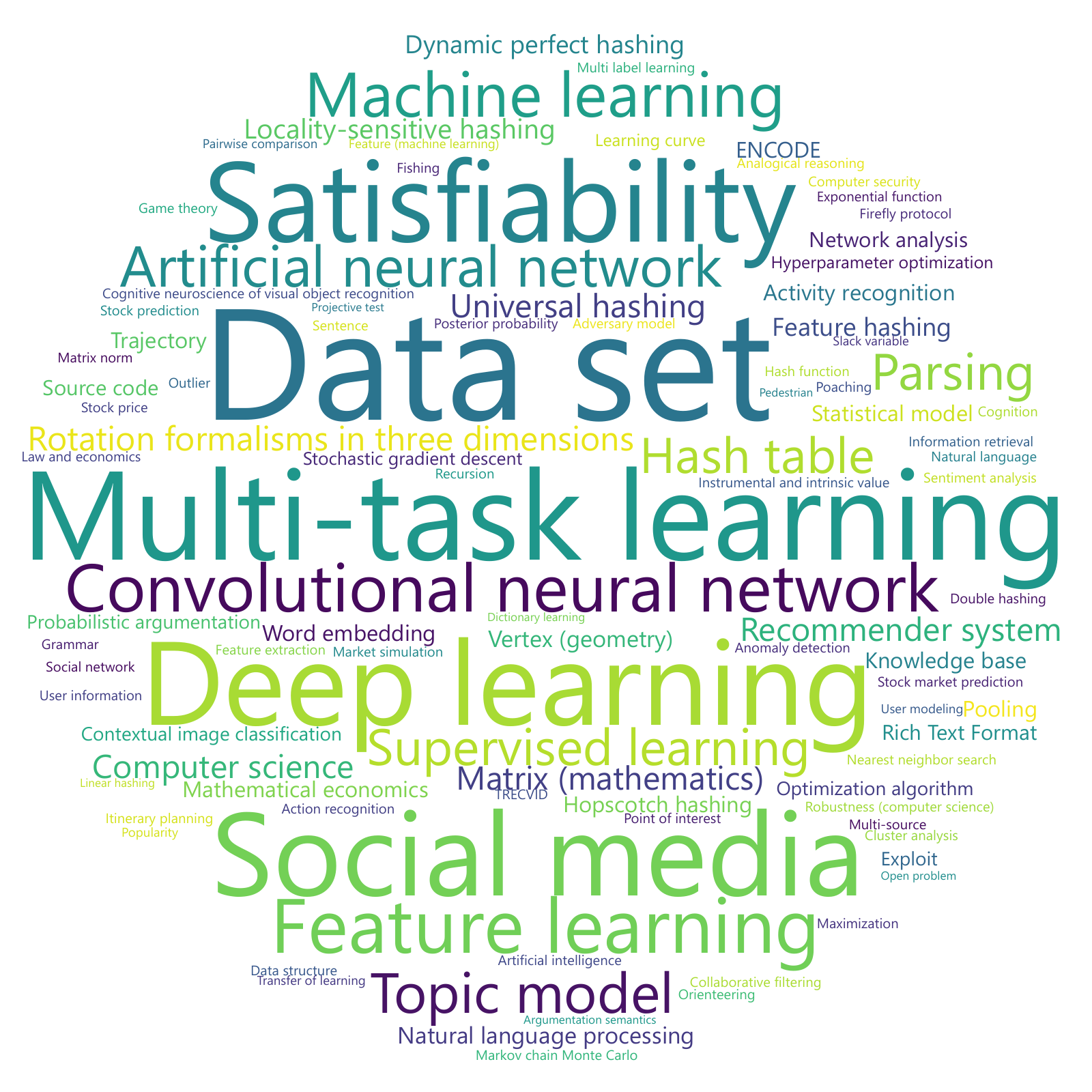}
    (b) IJCAI
 \end{minipage}
  \begin{minipage}{0.19\hsize}
  \centering
    \includegraphics[width=\hsize]{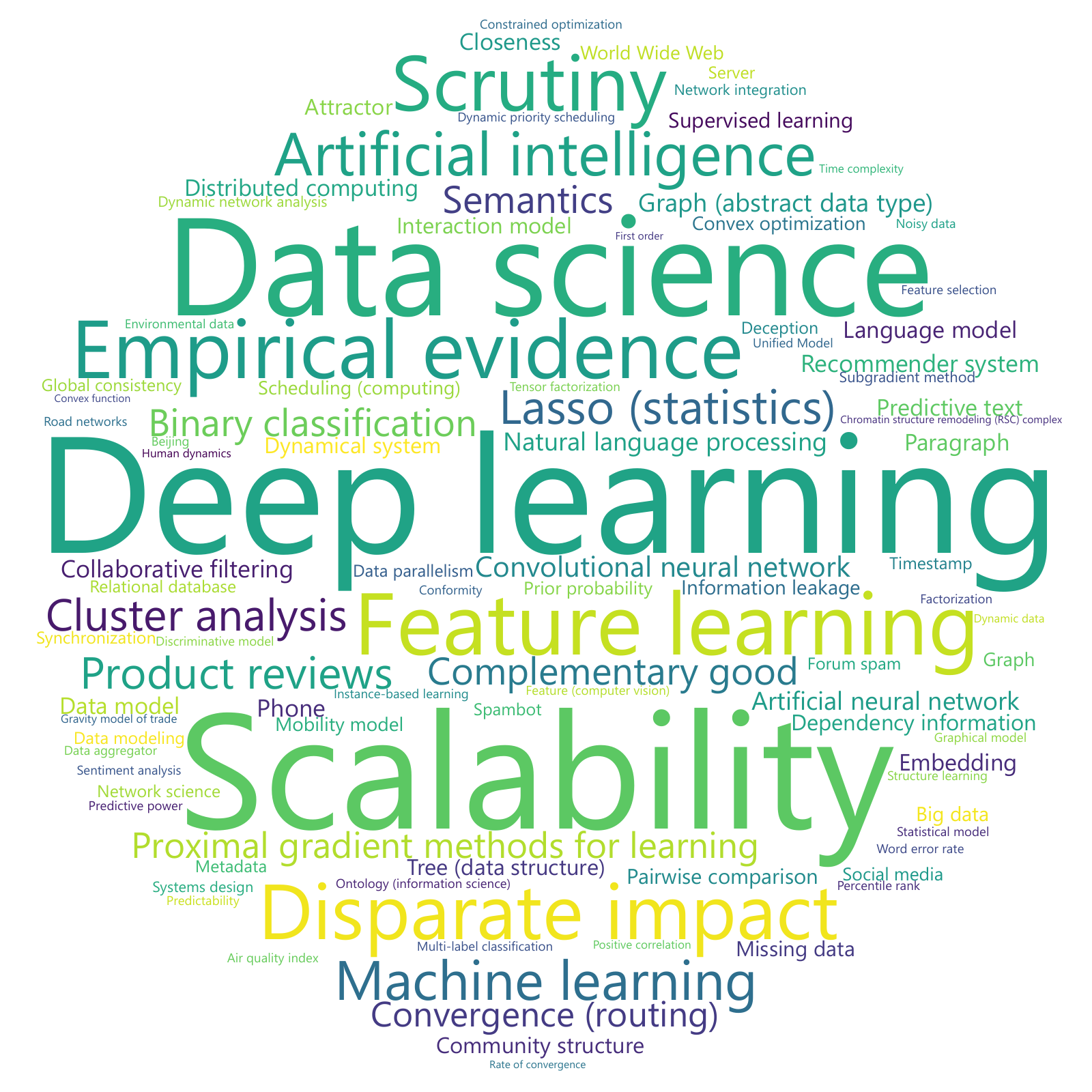}
    (c) KDD
 \end{minipage}
  \begin{minipage}{0.19\hsize}
  \centering
    \includegraphics[width=\hsize]{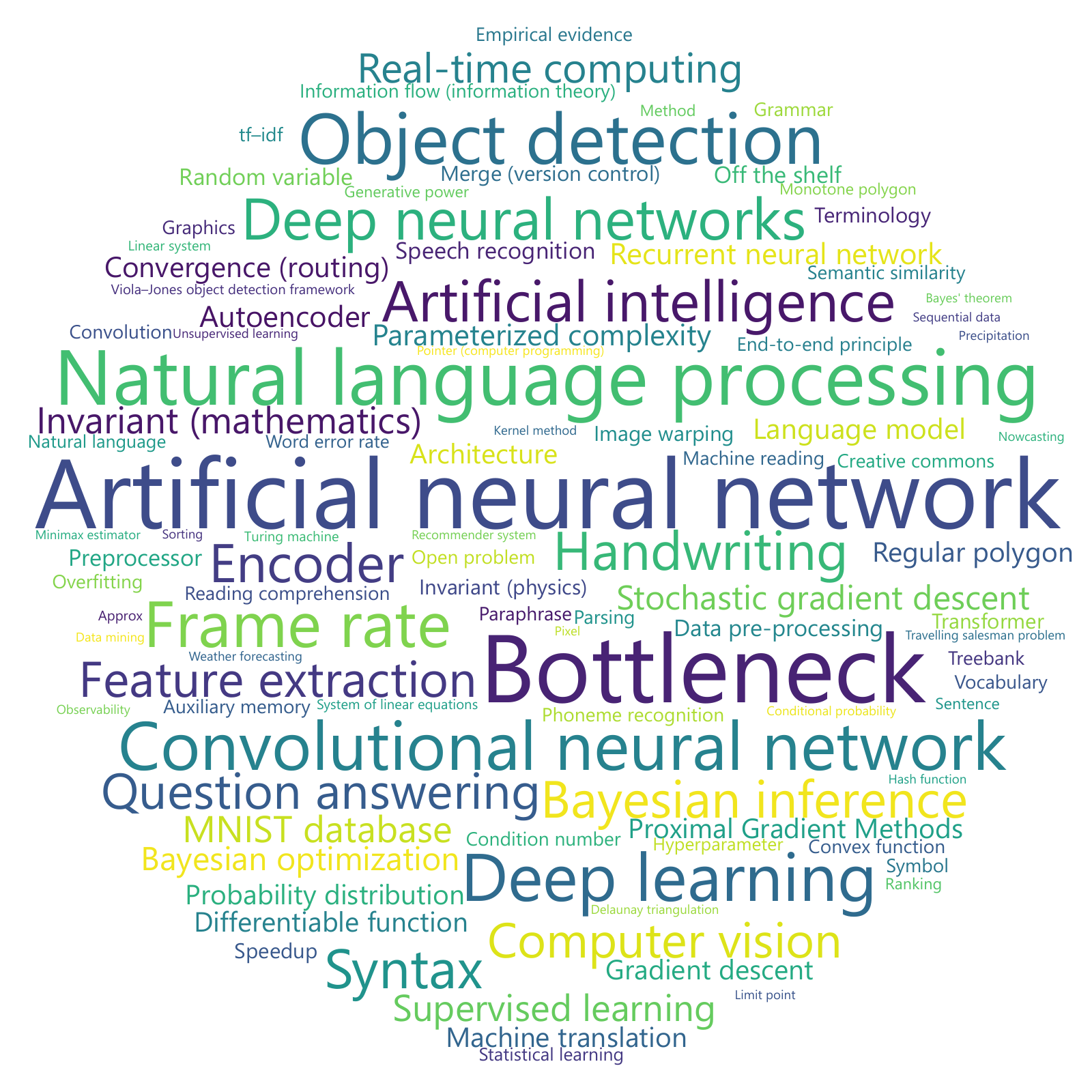}
    (d) NeurIPS
 \end{minipage}
  \begin{minipage}{0.19\hsize}
  \centering
    \includegraphics[width=\hsize]{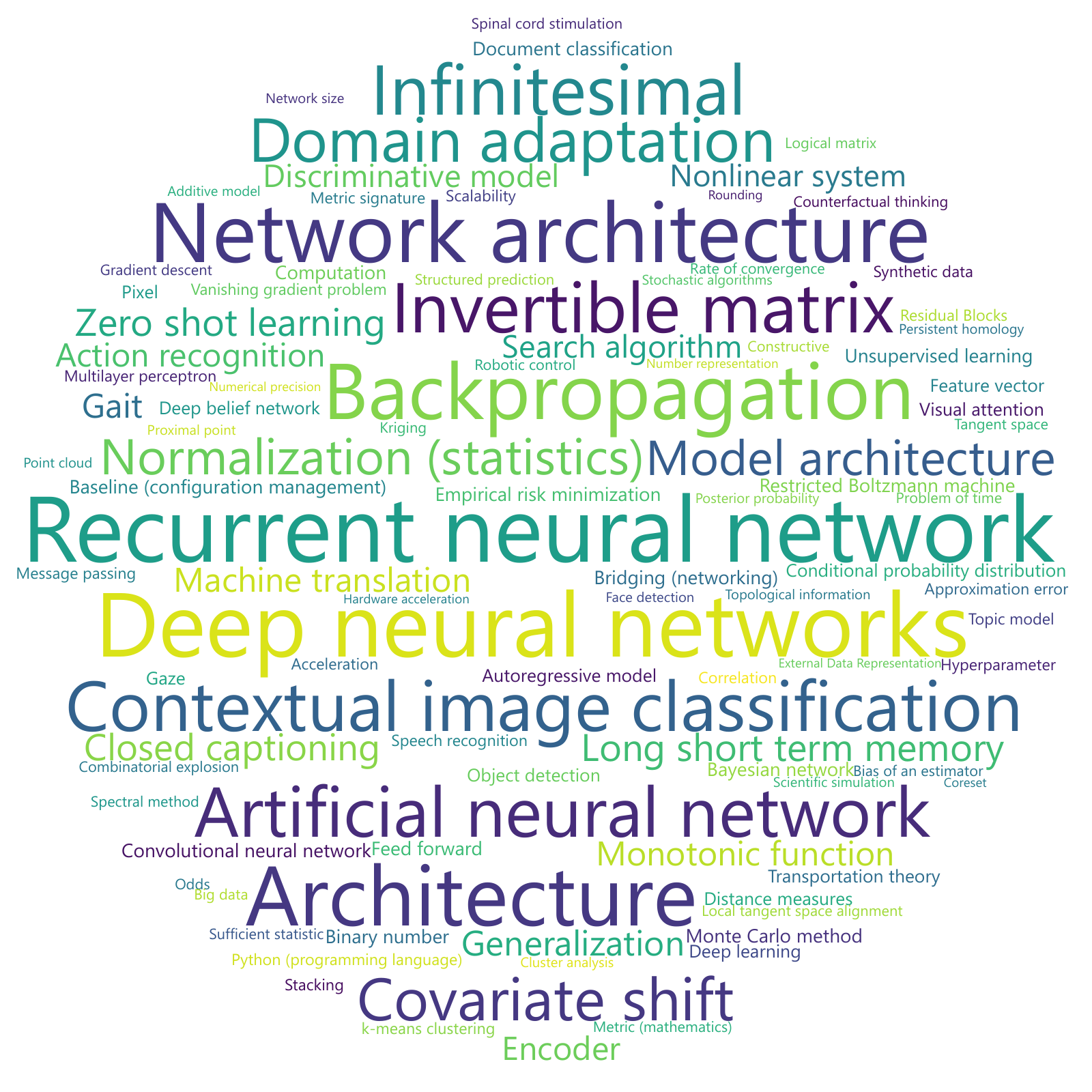}
    (e) ICML
 \end{minipage}
\caption{Interpretability: Important categories detected by \textsc{Poincare}. Categories with higher importance are written in larger fonts. Best viewed on screen.}
 \label{fig: interpretability}
\end{figure*}

In this section, we investigate the interpretability of \textsc{Poincare} via assessing the coefficients of the base learners $\hat{\mu}_t$. The larger coefficients of $\hat{\mu}_t$ indicate the importance of the topic for making a significant influence in venue $t$. Figure \ref{fig: interpretability} shows word clouds\footnote{\url{https://github.com/amueller/word_cloud}} with weights being the coefficients of the base learners of \textsc{Poincare} trained with the dblp dataset.
First, we can see that ``deep learning'' and ``deep neural networks'' have high importance in all conferences. This indicates that deep learning papers tend to gain many citations in all conferences.
Secondly, we can also see that different conferences focus on different topics.
For example, ``scalability'' is preferred in KDD ((c) bottom center). This is reasonable because extremely large data are often used in the data mining community.
``Network architecture'' ((e) top center) and ``architecture'' ((e) bottom center) are preferred in ICML. This indicates that proposing a new architecture of neural networks is influential in the ICML community.
We can also see that social media-related topics such as ``Social media'' ((b) bottom center, (a) bottom center) and ``Social network'' ((a) top center) have a high impact in artificial intelligence conferences. Sentiment analysis in Twitter \citep{you015robust} and estimating the demographics of social media \citep{culotta2015predicting} are examples of these topics. These papers are considered to be influential because many researchers in other fields, such as natural language processing researchers and physiologists, read AAAI and IJCAI papers, and they use and cite the methods introduced in artificial intelligence conferences.  We can validate such a tendency by enumerating papers that actually cite these papers. The interpretability of \textsc{Poincare} is reinforced by such post hoc analysis.
In addition to the benefits in decision making, the variety of important topics in different conferences indicates the different laws of outcomes in different venues, and it supports the choice of the T-learner over the S-learner.

\begin{description}
    \item[Observation 8.] \textsc{Poincare} reveals the preferred topics in each venue.
    \item[Observation 9.] Important topics are different in different venues.
\end{description}

\section{Limitation and Future Work}

\subsection{Other Treatment Effect Estimation Methods} Our proposed method utilizes only a propensity-based bias correction method. The experimental results suggest that other treatment effect estimation methods may also be useful for the publication venue recommendation problem. Exploring other treatment effect estimation methods, such as matching and hidden representation balancing, for venue recommendation is a promising future direction.

\subsection{Other Covariates} We used fields of study as covariates in the experiments. Our proposed framework is not limited to the particular choice of the covariates. There are other candidates for the covariates such as authors, countries, funds, and abstract text. In general, using more covariates leads to more accurate prediction and make the ignorability assumption more plausible. Investigating more covariates is important future work.

\subsection{Modeling Acceptance Probability} As we pointed out in Section \ref{sec: formulation}, modeling acceptance probability is an important future direction. Although we observed that \textsc{Poincare}'s recommendations resembled influential researchers' decision-making in Section \ref{sec: agreement}, it may be partially because non-influential researchers have limited choice of publication venues. Note that we selected all candidate venues, i.e., AAAI, IJCAI, KDD, NeurIPS, and ICML, from prestigious venues in each field and selected all candidate researchers from the last authors who have publications in these prestigious venues. Thus, we consider the effect of publication ability is less biased, and we have not observed any evidence that corroborates this concern. However, it would be an important direction to investigate how broad the option of each researcher is. Modeling acceptance probability would be an effective approach for this problem.

\section{Conclusion} We proposed a new formulation of the publication venue recommendation problem and proposed a recommender system that estimates the treatment effects of choosing a publication venue using a treatment effect estimation method. Unlike association-based recommender systems, our proposed method can recommend venues wherein the paper would have the most significant impact. Specifically, if the topic of paper A is a majority in venue X and is a minority in venue Y, an association-based recommender system always recommends venue X to paper A. This means that association-based recommender systems are biased towards historical records. By contrast, our proposed approach removes this bias and focuses on estimating how much impact the research has in each venue. Thereby, our proposed method may discover venue Y has many potential audiences of paper A, a thing that association-based recommender systems cannot do. Note that our method may also find publishing in familiar venues is the most effective for some papers, in which case our method agrees with association-based recommender systems.
We confirmed that the scores estimated by our approach correlated with the impacts of papers whereas the association-based approaches did not. We also confirmed that our method resembled influential researchers' decision making. These results indicate the advantages of the treatment effect-based approaches over the association-based approaches.

\section*{Acknowledgments}

This work was supported by the JSPS KAKENHI Grant Number 20H04243, 21J22490, and JST CREST Grant Number JPMJCR21D1.

\bibliography{citation}

\end{document}